\documentclass{article}
\usepackage{emulateapj,pstricks,apjfonts}

\def\asca       {{\em ASCA}\/}
\def\chandra    {{\em Chandra}\/}
\def\xmm 	{{\em XMM}\/}
\def\einstein   {{\em Einstein}\/}
\def\rosat      {{\em ROSAT}\/}
\def\am         {$^\prime$}

\def\cmsq       {~cm$^{-2}$}
\def\kms        {~km$\;$s$^{-1}$}
\def\msun	{~$M_{\odot}$}
\def\msunyr	{$M_{\odot}\;$yr$^{-1}$}
\def\lesssim{\mathrel{\hbox{\rlap{\hbox{\lower4pt\hbox{$\sim$}}}\hbox{$<$}}}}
\def\gtrsim{\mathrel{\hbox{\rlap{\hbox{\lower4pt\hbox{$\sim$}}}\hbox{$>$}}}}
\def\lax{\lesssim}
\def\gax{\gtrsim}
\def\hfifty	{$H_0$=50~km$\;$s$^{-1}\,$Mpc$^{-1}$}

\begin{document}

\lefthead{MASSES OF A2199 AND A496} 
\righthead{MARKEVITCH ET AL.}

\submitted{Submitted to ApJ}

\title{MASS PROFILES OF THE TYPICAL RELAXED GALAXY CLUSTERS A2199 AND A496}

\author{M. Markevitch\altaffilmark{1,2}, A. Vikhlinin\altaffilmark{1,2}, W.
R. Forman\altaffilmark{1}, and C. L. Sarazin\altaffilmark{3}}

\altaffiltext{1}{Harvard-Smithsonian Center for Astrophysics, 60 Garden St.,
Cambridge, MA 02138; maxim, alexey, wrf @head-cfa.harvard.edu}

\altaffiltext{2}{Space Research Institute, Russian Academy of Sciences}

\altaffiltext{3}{Astronomy Department, University of Virginia,
Charlottesville, VA 22903; cls7i@virginia.edu}

\begin{abstract}

We present maps and radial profiles of the gas temperature in the nearby
galaxy clusters A2199 and A496, which have the most accurate \asca\ spectral
data for all hot clusters. X-ray images, temperature maps, and the presence
of moderate cooling flows indicate that these clusters are relaxed and
therefore can provide reliable X-ray mass measurements under the assumption
of hydrostatic equilibrium and thermal pressure support. The cluster average
temperatures corrected for the presence of cooling flows are $4.8\pm0.2$ keV
and $4.7\pm0.2$ keV (90\% errors), respectively, which are 10\% higher than
the wide-beam single temperature fits.
Outside the central cooling flow regions and
within $r\approx 0.7\,h^{-1}$ Mpc covered by \asca, the radial temperature
profiles are similar to those of the majority of nearby relaxed clusters.
They are accurately described by polytropic models with $\gamma=
1.17\pm0.07$ for A2199 and $\gamma= 1.24_{-0.11}^{+0.08}$ for A496. We use
these polytropic models to derive accurate total mass profiles. Within
$r=0.5\,h^{-1}$ Mpc, which corresponds to a radius of overdensity 1000,
$r_{1000}$, for these clusters (estimated from our mass profiles), the total
mass values are $1.45\pm0.15\times 10^{14}\,h^{-1}$\msun\ and
$1.55\pm0.15\times 10^{14}\,h^{-1}$\msun. These values are 10\% lower than
those obtained assuming constant temperature. On the other hand, the values
inside a gas core radius ($0.07-0.13\,h^{-1}$ Mpc) are a factor of $\gax
1.5$ higher than the isothermal values. The gas mass fraction increases
significantly with radius (by a factor of 3 between the X-ray core radius
and $r_{1000}$) and at $r_{1000}$ reaches similar values of $0.057\pm0.005\,
h^{-3/2}$ and $0.056\pm0.006\, h^{-3/2}$ for the two clusters, respectively.
Our measured total mass profiles within $r_{1000}$ are remarkably well
approximated by the Navarro, Frenk, and White ``universal'' profile. Since
A2199 and A496 are typical relaxed clusters, the above findings should be
relevant for most such systems. In particular, the similarity of the
temperature profiles in nearby clusters appears to reflect the underlying
``universal'' dark matter profile. The upward revision of the mass values at
small radii for the observed temperature profile compared to those derived
assuming isothermality will resolve most of the discrepancy between the
X-ray and strong lensing mass estimates.

\end{abstract}

\keywords{cooling flows --- dark matter --- galaxies: clusters: individual
(A496, A2199) --- intergalactic medium --- X-rays: galaxies}

\section{INTRODUCTION}

\begin{table*}[tb]
\small
\renewcommand{\arraystretch}{1.3}
\renewcommand{\tabcolsep}{2mm}
\begin{center}
TABLE 1

{\sc Cluster Parameters$^{\rm a}$}
\vspace{1mm}

\begin{tabular}{p{1.8cm}cccccccccc}
\hline \hline
Cluster& $a_x^{\rm b}$ & $\beta^{\rm b}$ &$\rho_{\rm gas,0}^{\rm b}$ %
  & $T_e^{\rm c}$ & $T_X^{\rm d}$ %
  & $M(0.2\;{\rm Mpc})$ & $M(1\;{\rm  Mpc})$ & $f_{\rm gas}(1\;{\rm  Mpc})$ %
  & $M(r_{500})^{\rm e}$ & $r_{200}/r_s^{\rm ~~e}$ \\
       & kpc           &               & $M_\odot {\rm Mpc}^{-3}$  %
  & keV           & keV  %
  & $10^{14}$\msun        & $10^{14}$\msun    &   & $10^{14}$\msun & \\
\hline
A2199 \dotfill&134 &0.636 & $2.24\times10^{14}$& $4.4\pm0.2$ &$4.8\pm0.2$ %
      & $0.65\pm0.11$ &$2.9\pm0.3$ & $0.161\pm 0.014$ & $3.6\pm0.5$ & 10 \\
A496 \dotfill &249 &0.700 & $1.02\times10^{14}$& $4.3\pm0.2$ &$4.7\pm0.2$ %
      & $0.47\pm0.10$ &$3.1\pm0.3$ & $0.158\pm 0.017$ & $3.9\pm0.6$ & 6  \\
\hline
\end{tabular}
\vspace{1mm}

\begin{minipage}{17.5cm}
$^{\rm a}$All values are for $h=0.5$. $^{\rm b}$Best-fit values for the
\rosat\ PSPC brightness profile excluding the central $r=3'$. The gas
density value is an extrapolation of this $\beta$-model to the center.
$^{\rm c}$Wide-beam single temperature fit. $^{\rm d}$Emission-weighted
temperature excluding cooling flows. $^{\rm e}$Involves extrapolation to
the area not covered by the temperature profile.
\end{minipage}
\end{center}
\vspace*{-2mm}
\end{table*}

Under a reasonable, but as yet not directly tested, set of assumptions that
the hot intracluster gas is supported by its own thermal pressure and is in
hydrostatic equilibrium in the cluster gravitational well, one can determine
the total mass of a cluster, including its dominant dark matter component
(Bahcall \& Sarazin 1977; Mathews 1978). Because clusters are the largest
collapsed objects in the Universe, their mass values are of great importance
for cosmology. The cluster mass function and its evolution with redshift
constrain the spectrum of the cosmological density fluctuations and the
density parameter $\Omega_0$ (e.g., Press \& Schechter 1974; White \& Rees
1978; Bahcall \& Cen 1992; Viana \& Liddle 1996). If the cluster matter
inventory is representative of the Universe as a whole, as is expected, then
by measuring the cluster total and baryonic mass and comparing it to the
predictions of primordial nucleosynthesis, one can constrain $\Omega_0$
(White et al.\ 1993). Comparison of independent cluster mass estimates, for
example, by X-ray and gravitational lensing (e.g., Bartelmann \& Narayan
1995) methods, provide unique insights into cluster structure and physics. A
discrepancy between the different estimates may indicate significant
turbulence or nonthermal pressure in the intracluster gas (e.g., Loeb \& Mao
1994), or the effect of line of sight projections.

For an X-ray measurement of the cluster mass, one needs accurate radial
profiles of the gas density and temperature, as well as confidence that the
cluster is in hydrostatic equilibrium. The gas density profile for a symmetric
cluster can readily be obtained with an imaging instrument, such as
\einstein\ or \rosat. Obtaining temperature distributions has proven to be
more problematic, especially for hotter, more massive clusters.
\asca\ (Tanaka, Inoue, \& Holt 1994) now provides spatially resolved
temperature data for nearby hot clusters (e.g., Ikebe et al.\ 1997;
Loewenstein 1997; Donnelly et al.\ 1998; Markevitch et al.\ 1998 [hereafter
MFSV] and references therein), although their accuracy is still limited.
Outside the central cooling flow regions, the temperature decreases with
radius in most studied clusters. For a few clusters with more accurate
temperature profiles, accurate mass profiles were already obtained (e.g.,
for A2256 by Markevitch \& Vikhlinin 1997, hereafter MV).

MFSV found that gas temperature profiles of nearby symmetric clusters
outside the cooling flow regions are similar when scaled by the virial
radius and average temperature. The gas density profiles also are rather
similar (e.g., Jones \& Forman 1984; Vikhlinin, Forman, \& Jones 1999). This
suggests that the underlying dark matter profiles are similar. Indeed,
analytical work and cosmological cluster simulations (e.g., Bertschinger
1985; Cole \& Lacey 1996; Navarro, Frenk, \& White 1995, 1997,
hereafter NFW) predict that the dark matter radial profiles of most clusters
in equilibrium should be similar in units of the virial radius. It is
interesting to see whether their predicted ``universal'' dark matter profile
agrees with the observations.

For all but a few clusters in the MFSV sample, the temperature data have
insufficient accuracy for such a test. We therefore selected two additional
typical, relaxed, but less distant clusters, A2199 ($z=0.030$) and A496
($z=0.033$), for a more accurate temperature profile and mass derivation
using \asca. These clusters are very ordinary in their X-ray luminosities
and temperatures ($T\simeq 4.5$ keV) and, similarly to most clusters, have
moderate cooling flows (170 and 95 \msunyr, respectively; Peres et al.\
1998). The presence of cooling flows is suggestive of a relaxed cluster,
while at the same time these flows are not so strong as to prevent accurate
resolved temperature measurements with ASCA (see MFSV).  A subset of the
data presented here (observations of the central regions) was already
analyzed by Mushotzky et al.\ (1995). We have since obtained offset
observations, and include the \asca\ PSF correction in our analysis. Below
we use \asca\ and \rosat\ data on these two clusters to derive their total
mass profiles. We use \hfifty\ ($h=0.5$); the error intervals are 90\%.

\begin{figure*}[tb]
\pspicture(0,9.8)(18.5,22.9)

\rput[tl]{0}(0.5,23.2){\epsfxsize=8.5cm
\epsffile{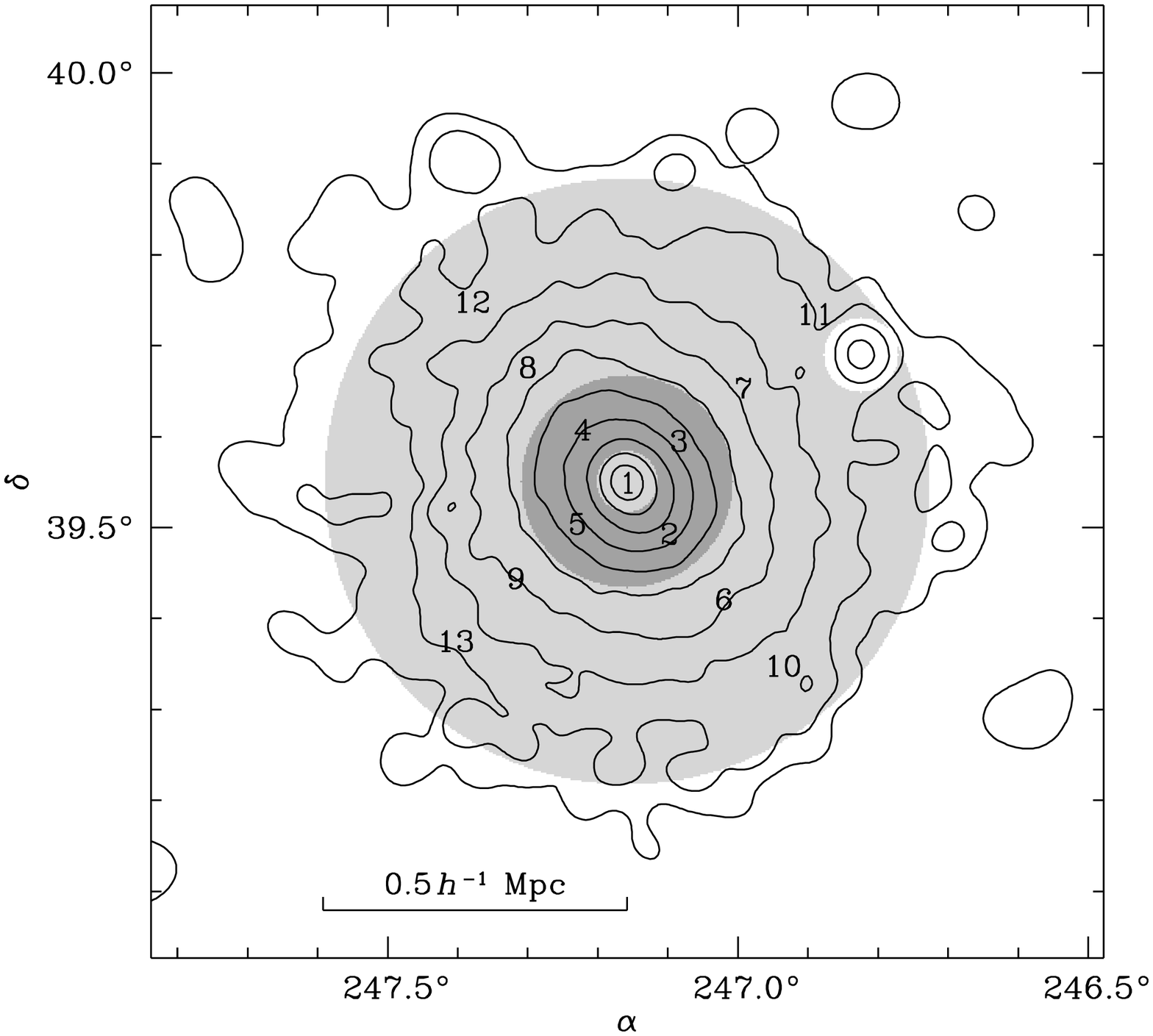}}

\rput[tl]{0}(0.75,16.2){\epsfxsize=7.75cm
\epsffile[30 470 530 678]{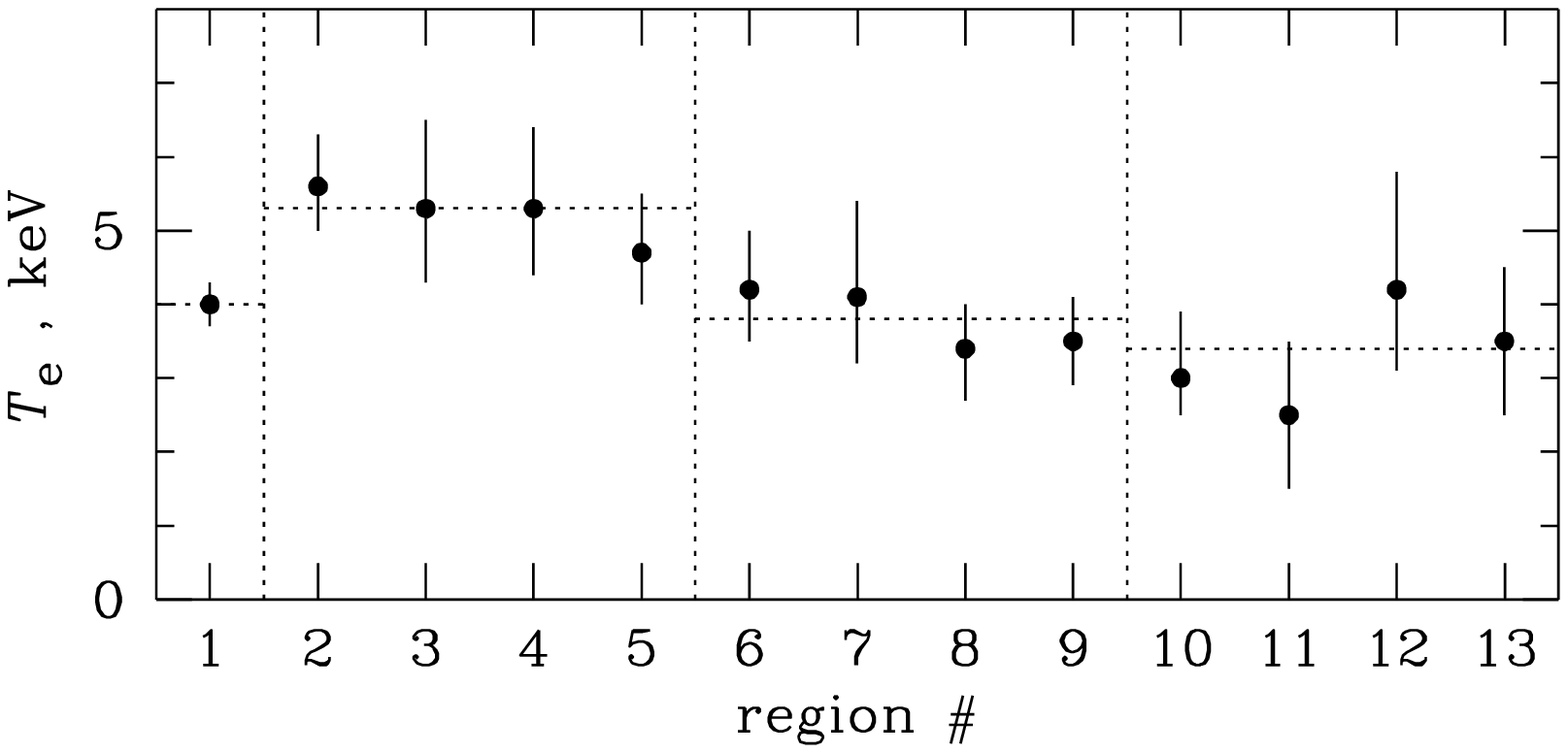}}

\rput[tl]{0}(9.5,23.2){\epsfxsize=8.5cm
\epsffile{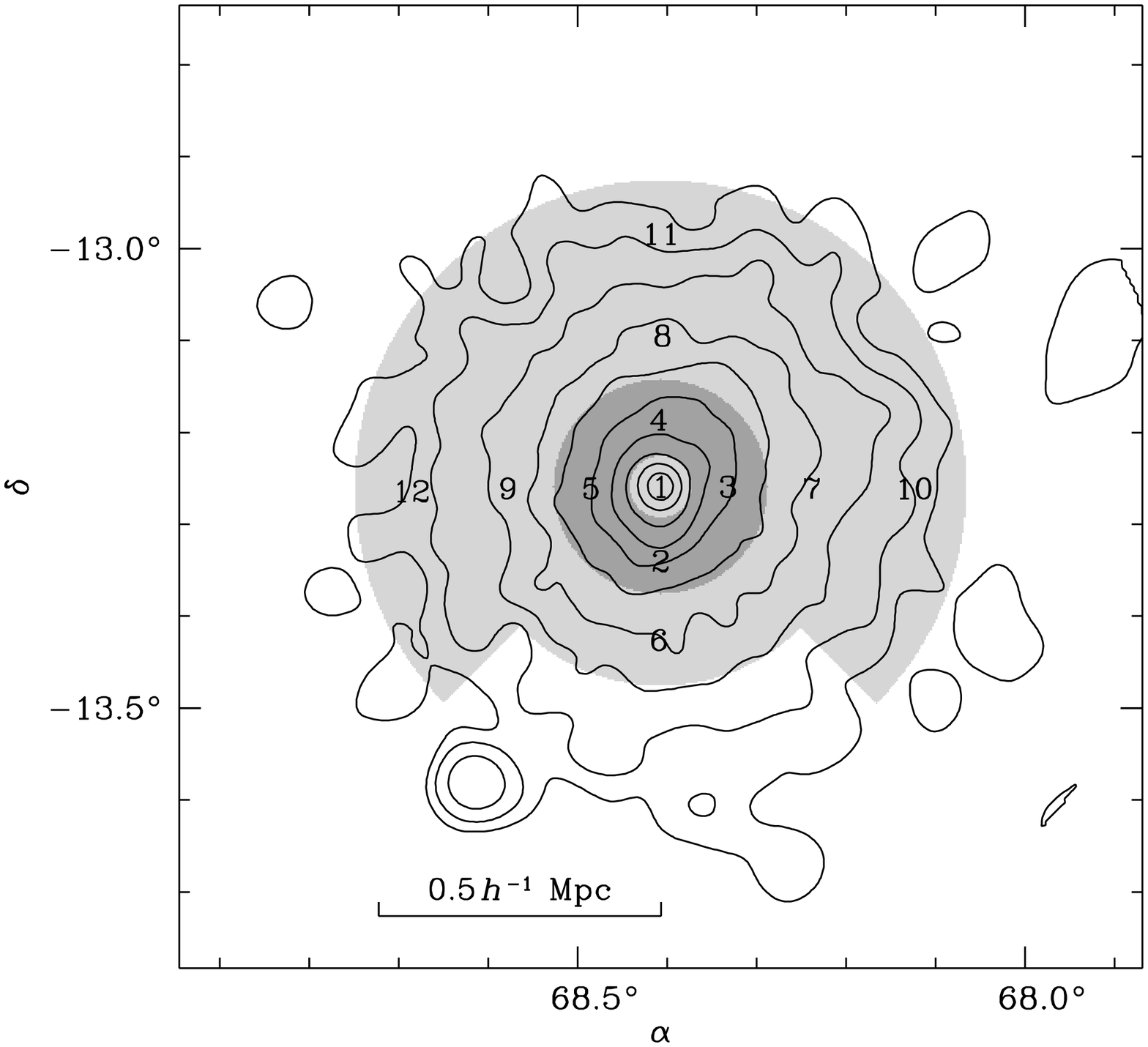}}

\rput[tl]{0}(9.75,16.2){\epsfxsize=7.75cm
\epsffile[30 470 530 678]{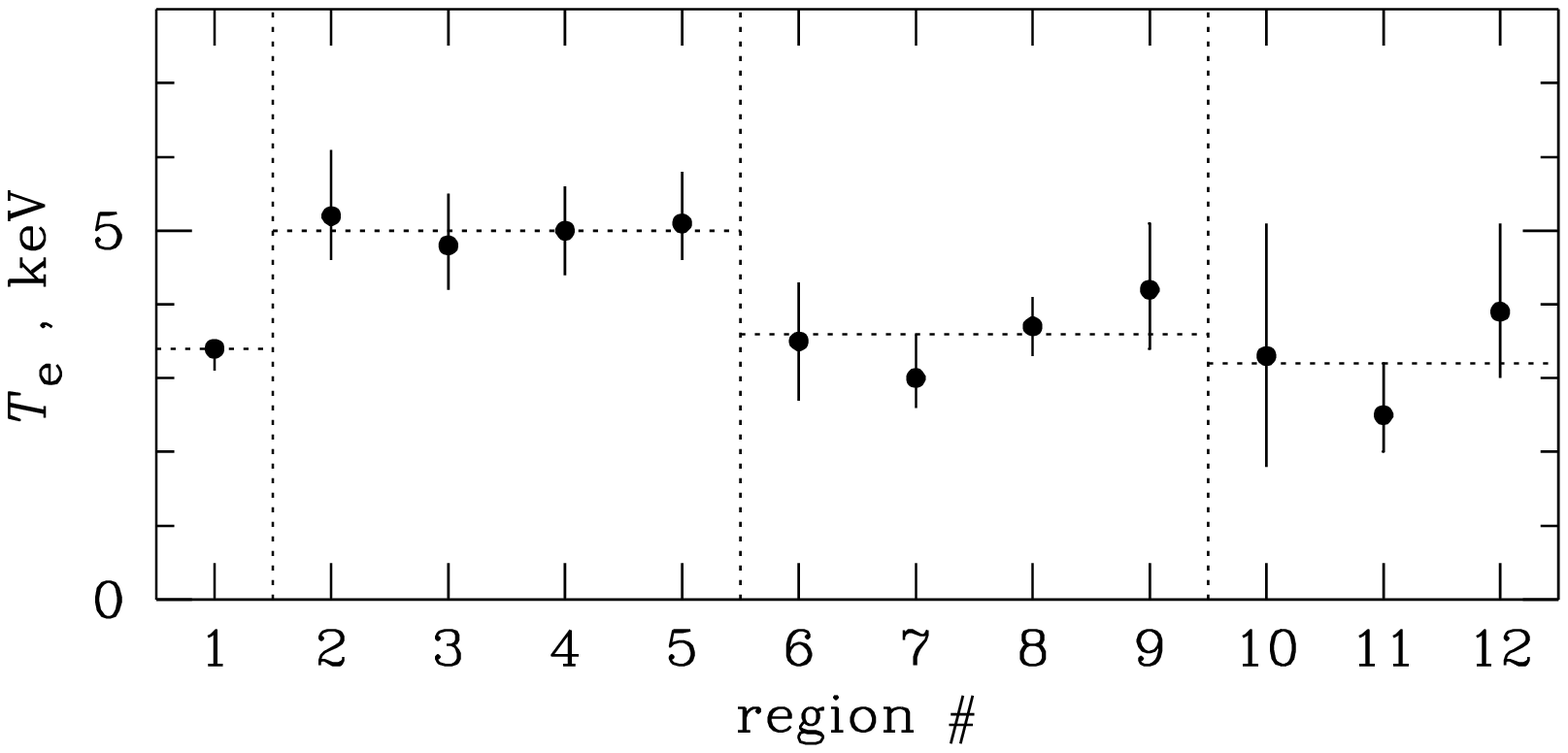}}

\rput[l]{0}(2.3,22.5){\small A2199}
\rput[l]{0}(11.3,22.5){\small A496}

\rput[tl]{0}(0,12.0){
\begin{minipage}{18cm}
\small\parindent=3.5mm
{\sc Fig.}~1.---\asca\ projected temperature maps (color) overlaid on the
\rosat\ PSPC brightness contours spaced by a factor of 2. Sectors in which
the temperature was derived are numbered in upper panels and the
temperatures with 90\% errors are given in lower panels. Different colors
correspond to significantly different temperatures. Vertical dotted lines
separate regions belonging to different annuli. Dotted horizontal lines show
average temperature inside the annulus. For the central cooling flow
regions, a single-temperature fit is shown. A white circle in the map shows a
point source excluded from the fit.
\end{minipage}
}
\endpspicture
\end{figure*}

\begin{figure*}[tb]
\pspicture(0,10.2)(18.5,21)

\rput[tl]{0}(0.4,20.7){\epsfxsize=8.5cm
\epsffile{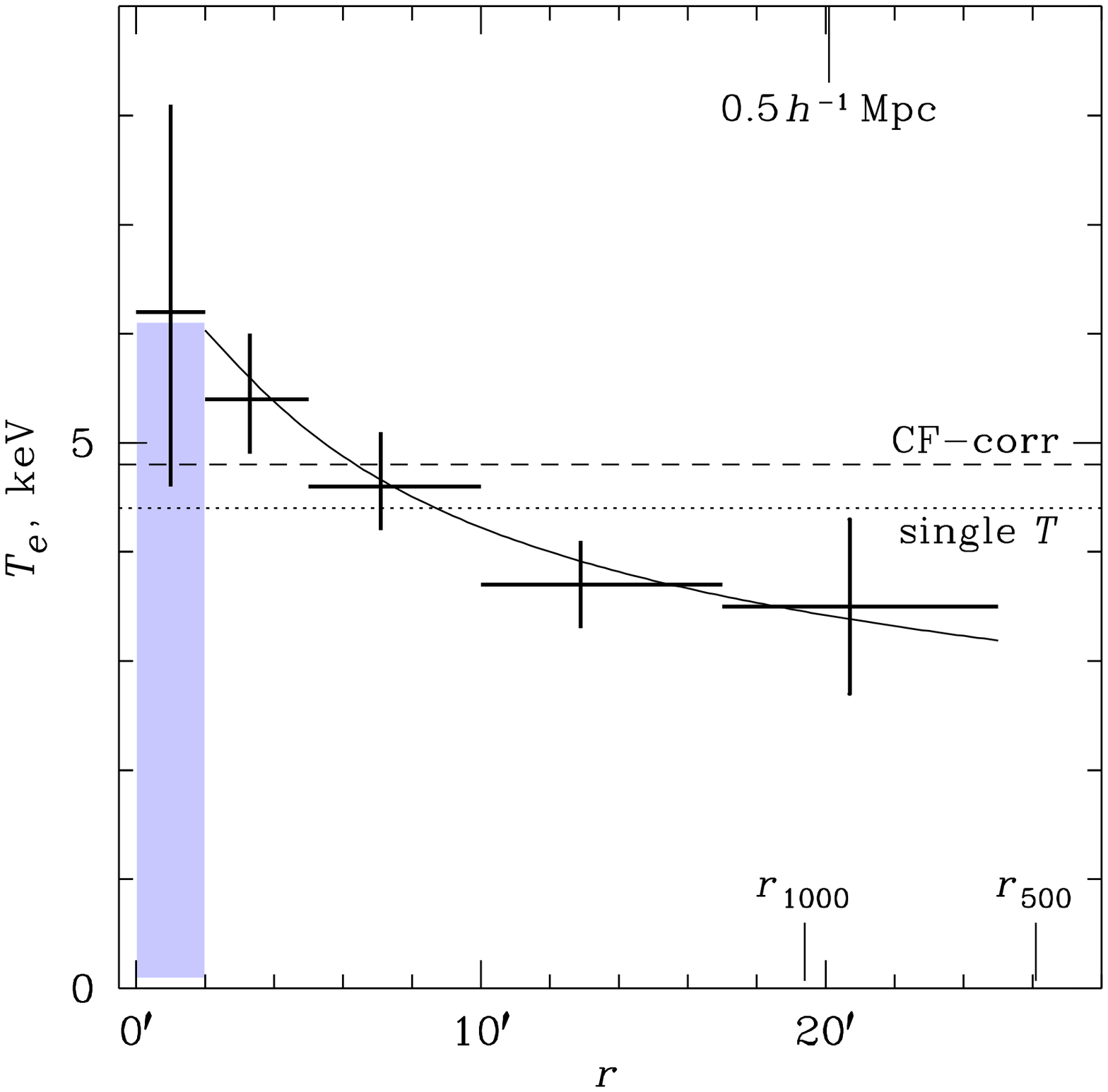}}

\rput[tl]{0}(9.5,20.7){\epsfxsize=8.5cm
\epsffile{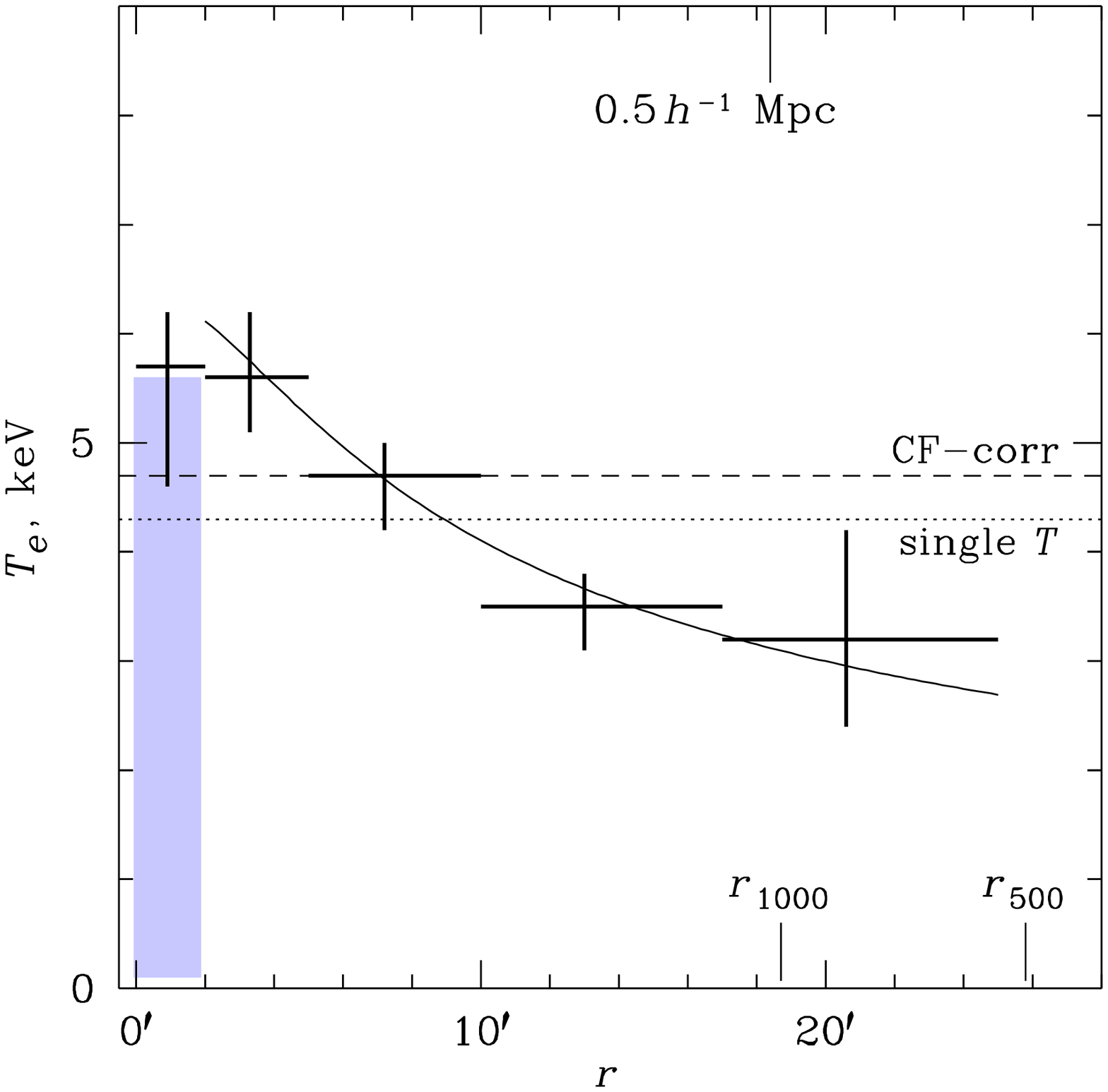}}

\rput[bl]{0}( 2.3,19.4){\small A2199}
\rput[bl]{0}(11.4,19.4){\small A496}

\rput[tl]{0}(0,12.4){
\begin{minipage}{18cm}
\small\parindent=3.5mm
{\sc Fig.}~2.---Radial projected temperature profiles. Crosses are centered
on the emission-weighted radii. Vertical errors are for 90\% confidence;
horizontal error bars show the boundaries of the annulus. Gray bands denote
a continuous range of temperatures in a cooling flow, and the central cross
corresponds to the upper (ambient) temperature of the cooling flow. Dotted
and dashed horizontal lines show wide beam single-temperature fits and
average emission-weighted temperatures excluding the cooling flow. Smooth
lines show polytropic fits to the values outside the central bin,
$\gamma=1.17$ and 1.24 for A2199 and A496, respectively. Values of
$r_{1000}$ and $r_{500}$ are calculated from the mass profiles obtained from
these temperature data (see Figs.\ 5 and 6 below).
\end{minipage}
}
\endpspicture
\end{figure*}

\section{\rosat\ PSPC DATA}
\label{rossec}

To derive the gas density distribution, we use \rosat\ PSPC data. The
archival observations of A2199 and A496 were analyzed as prescribed by
Snowden et al.\ (1994) and using S. Snowden's code. To optimize the signal
to noise ratio, we used Snowden bands 5--7 that correspond to 0.7--2.0
keV. For A2199, two observations of the same field were combined.  The
radial brightness profiles were then fit with a $\beta$-model $S_X(r)
\propto (1+r^2/a_x^2)^{-3\beta+\frac{1}{2}}$ plus a uniform X-ray background
within a radial range of $3'-50'$. The inner radius of 3\am\ approximately
corresponds to the cooling radius for both clusters (e.g., Peres et al.\
1998) and encompasses all of the X-ray brightness excess due to the moderate
cooling flows in the cluster centers. The resulting parameters of the gas
density profile are given in Table 1 and are typical (e.g., Jones \& Forman
1984). The $\beta$-model values for A2199 are similar to the results of
Siddiqui, Stewart, \& Johnstone (1998) using the same data.

\section{\asca\ DATA}

A2199 and A496 were each observed by \asca\ with one central pointing and
two different 14--15\am\ offsets from the cluster centers. Such a
configuration has been chosen to cover the cluster to a radius where the
mean overdensity is 500 ($r_{500}$), while at the same time keeping the
cluster brightness peak within the \asca\ field of view to avoid stray light
contamination. Observing the clusters at different positions in the focal
plane also reduces the \asca\ systematic uncertainties that dominate in the
temperature estimates. The offset positions were chosen to avoid bright
foreground sources and also to cover representative regions of these
slightly elliptical clusters.

After the standard data screening (ABC Guide%
\footnote{http://heasarc.gsfc.nasa.gov/docs/asca/abc/abc.html}%
), useful GIS exposures for the A2199 central and offset pointings were 31
ks, 19 ks, and 22 ks, and for A496, they were 37 ks, 24 ks, and 24 ks,
respectively (the corresponding SIS exposures were about a factor of 0.8 of
the GIS exposures). For the temperature fits, all pointings for both GIS and
SIS were used simultaneously; different pointings and instruments fitted
separately give consistent results. To derive the spatial temperature
distributions, we used the method described in detail in MFSV and references
therein. This method accounts for the \asca\ PSF and assumes that outside
the cooling flow regions, the \rosat\ PSPC image provides an accurate
description of the relative spatial distribution of the projected gas
emission measure, after a correction of the PSPC brightness for any gas
temperature variations. It should be mentioned here that a recent discovery
of the possibly nonthermal EUV and soft X-ray ($E<0.2$ keV) emission in
A2199 should not affect the latter assumption in any significant way, since
we use a relatively hard ($0.7-2.0$ keV) PSPC band where this excess is
absent (Lieu, Bonamente, \& Mittaz 1999).  The absorption column was assumed
uniform at the Galactic values ($N_H=0.9\times 10^{20}$\cmsq\ and $4.6\times
10^{20}$\cmsq\ for A2199 and A496); for our $E>1.5$ keV spectral fitting
band, any expected variations are unimportant.

The analysis method propagates all known calibration and other systematic
uncertainties, including those of the \asca\ PSF, effective area,
\rosat\ and \asca\ backgrounds etc., to the final temperature values. All
reported confidence intervals are one-parameter 90\% and are estimated by
Monte-Carlo simulations.

\section{RESULTS}

\subsection{Temperature Maps}

The resulting two-dimensional projected temperature maps are shown in Fig.\
1, overlaid on the \rosat\ images. We show only sectors in which the
temperature is accurately constrained. The maps show no significant
azimuthally asymmetric variations, and together with the brightness contours
suggest that these clusters are well relaxed. These maps may be contrasted
to the similarly derived, but highly irregular, temperature maps of merging
clusters, e.g., A754 (Henriksen \& Markevitch 1996) and Cygnus-A and A3667
(Markevitch, Sarazin, \& Vikhlinin 1999). In the central regions, the maps
clearly show low temperature regions that correspond to the previously known
cooling flows (e.g., Stewart et al.\ 1984; Edge, Stewart, \& Fabian 1992).

\subsection{Radial Temperature Profiles}
\label{tprof}

Figure 2 shows the cluster projected temperature profiles in five annuli.
For the central radial bin, we used a model consisting of a thermal
component and a cooling flow with the upper temperature tied to that of the
thermal component, both with free normalizations. The figure also shows
wide-beam, single-temperature fits ($T_e=4.4\pm0.2$ keV and $4.3\pm0.2$ keV
for A2199 and A496, respectively) and emission-weighted average temperatures
excluding the cooling flow component ($T_X=4.8\pm0.2$ keV and $4.7\pm0.2$
keV).  The latter are calculated from these temperature profiles as
described in MFSV. The data indicate a higher temperature in the central
cluster regions (outside the cooling flows) compared to the average
temperature, and a temperature decline with radius. This is similar to other
clusters; in fact, when the profiles for A2199 and A496 are plotted in units
of $T_X$ and virial radius, they lie within the composite profile obtained
by MFSV for other nearby, relatively symmetric clusters (Fig.\ 3). Such
typical temperature profiles, together with the typical gas density profiles
and the presence of cooling flows, make A2199 and A496 representative
examples of relaxed clusters.

Outside the central cooling flow bin, the profiles in Fig.\ 2 are described
remarkably well by a polytrope, $T_{\rm gas}\propto \rho_{\rm
gas}^{\gamma-1}$ (both temperature profiles appear slightly more concave
than the polytropic fits, which is probably nothing more than a coincidence;
note that they differ in a similar way from the composite profile in Fig.\
3). Assuming the \rosat-derived $\beta$-models for $\rho_{\rm gas}$, we find
$\gamma=1.17\pm0.07$ and $\gamma=1.24_{-0.11}^{+0.08}$ for A2199 and A496,
respectively. Regardless of whether this fact has any physical meaning or is
purely fortuitous, it simplifies the total mass derivation by providing a
convenient functional form for the observed temperature profile.  We will
use it in the next section, but first note that for the mass derivation, one
needs a real (three-dimensional) gas temperature profile as opposed to
projected on the plane of the sky that we have obtained. We show in Appendix
that as long as the gas density follows a $\beta$-model and the temperature
is proportional to a power of density, a projected polytropic temperature
profile differs from the three-dimensional profile only by a normalization.
For the best-fit $\beta$ and $\gamma$ values for A2199 and A496, the
projected temperature profiles are factors of 0.94 and 0.92 lower than the
three-dimensional profiles, respectively.

\section{TOTAL MASS PROFILES}
\label{masssec}

For the total mass determination, we will take advantage of the fact that
the temperature profiles can be described by a polytropic functional form.
>From the hydrostatic equilibrium equation for a spherically symmetric gas
distribution $\rho_{\rm gas}(r)\propto (1+r^2/a_x^2)^{-\frac{3}{2}\beta}$
and a temperature profile $T\propto \rho_{\rm gas}^{\gamma-1}$, the total
mass within a radius $r=xa_x$ is given by
\begin{equation}
\label{eq:m}
M(r) = 3.70\times 10^{13} M_\odot\, \frac{0.60}{\mu}\, \frac{T(r)}{\rm
1\;keV}\, \frac{a_x}{\rm 1\;Mpc}\, \frac{3 \beta \gamma x^3}{1+x^2}
\end{equation}
(see, e.g., Sarazin 1988). A polytropic temperature decline thus corresponds
to the following correction to an isothermal mass estimate $M_{\rm iso}$:

\pspicture(0,-1.8)(8.5,9.8)

\rput[tl]{0}(-0.5,9.7){\epsfxsize=8.8cm
\epsffile{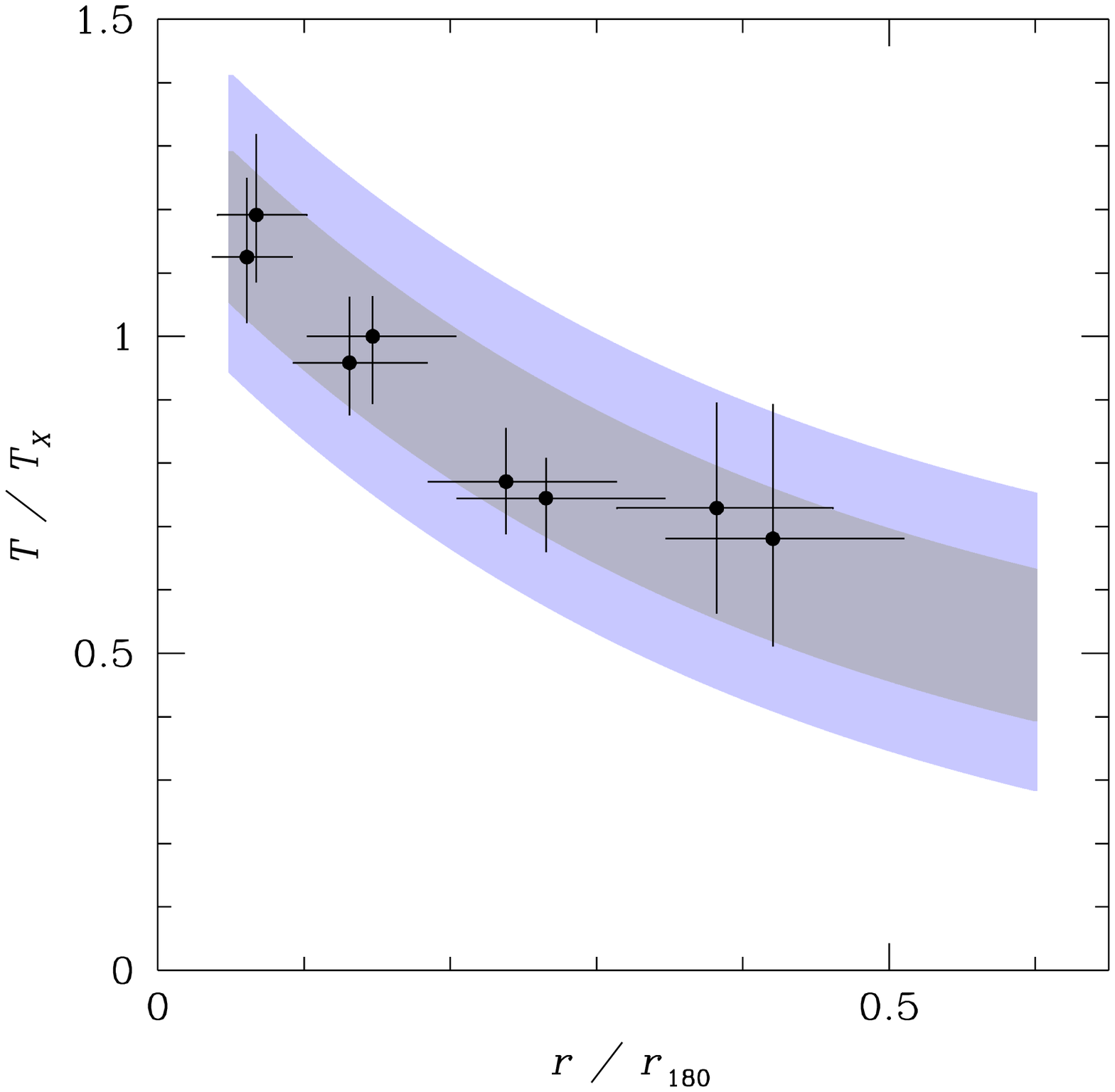}}

\rput[tl]{0}(-0.45,1){
\begin{minipage}{8.75cm}
\small\parindent=3.5mm
{\sc Fig.}~3.---Temperature profiles for A2199 and A496 (symbols) overlaid
on the gray band representing a composite profile for a sample of 19 nearby
relatively symmetric clusters from MFSV (their Figs.\ 7 and 8). For this
comparison, the profiles are normalized by their cooling flow-corrected
average temperatures $T_X$ and plotted in units of $r_{180}$ estimated from
$T_X$ using the relation of Evrard et al.\ (1996). Cooling flow bins are not
shown. The darker band corresponds to a scatter of best-fit temperature
values of the MFSV sample and the lighter band covers most of their 90\%
intervals.
\end{minipage}
}
\endpspicture

\begin{equation}
\label{eq:mcor}
\frac{M(r)}{M_{\rm iso}(r)}=\frac{T(r)}{\overline{T}}\,\gamma,
\end{equation}
where $\overline{T}$ is the average temperature (see also Ettori \& Fabian
1999). To calculate the 90\% confidence bands on mass profiles (as well as
the confidence intervals on the values of $\gamma$ above), we have fitted
the polytropic model to the same simulated temperature values in those
annuli that were used to calculate the temperature error bars (see MFSV).
These fitted polytropic models were substituted into equations (1) and (2)
above and 90\% confidence intervals of the resulting values were calculated
at each radius. The resulting correction factor to the isothermal mass
estimate is shown in Fig.\ 4 as a function of radius. The corresponding
profiles of the total mass, $M$, and the gas mass fraction, $f_{\rm
gas}\equiv M_{\rm gas}/M$, are shown in Fig.\ 5. The corresponding ratio of
the mean total density within a given radius to the critical density at the
cluster's redshift [$\rho_c=3 H_0^2 (1+z)^3 /8\pi G$] is shown as a function
of radius in Fig.\ 6. For both clusters, our mass profiles correspond to
$r_{1000}\approx 1.0$ Mpc and $r_{500}\approx 1.3-1.4$ Mpc (the latter
involves extrapolation to a region not covered by the temperature
profile, see Fig.\ 2). Masses and gas fractions at several interesting radii
are also given in Table 1.

At $r=1$ Mpc, Mushotzky et al.\ (1995) obtained mass estimates of
$2.55\times 10^{14}$\msun\ for A2199 and $3.05\times 10^{14}$\msun\ for
A496. These estimates are close to ours, even though Mushotzky et al.\ did
not apply the \asca\ PSF correction in their analysis. From the galaxy
velocity data, Girardi et al.\ (1998) obtained, at $r=1$ Mpc, masses of
$5.4^{+2.4}_{-1.9}\times 10^{14}$\msun\ for A2199 and $3.5\pm 1.8\times
10^{14}$\msun\ for A496 (their 68\% errors are multiplied by 1.65 to obtain
90\% intervals). These are consistent (although for A2199, only marginally)
with our more accurate values.

\begin{figure*}[tb]
\pspicture(0,11.2)(18.5,21)

\rput[tl]{0}(0.4,20.7){\epsfxsize=8.5cm
\epsffile{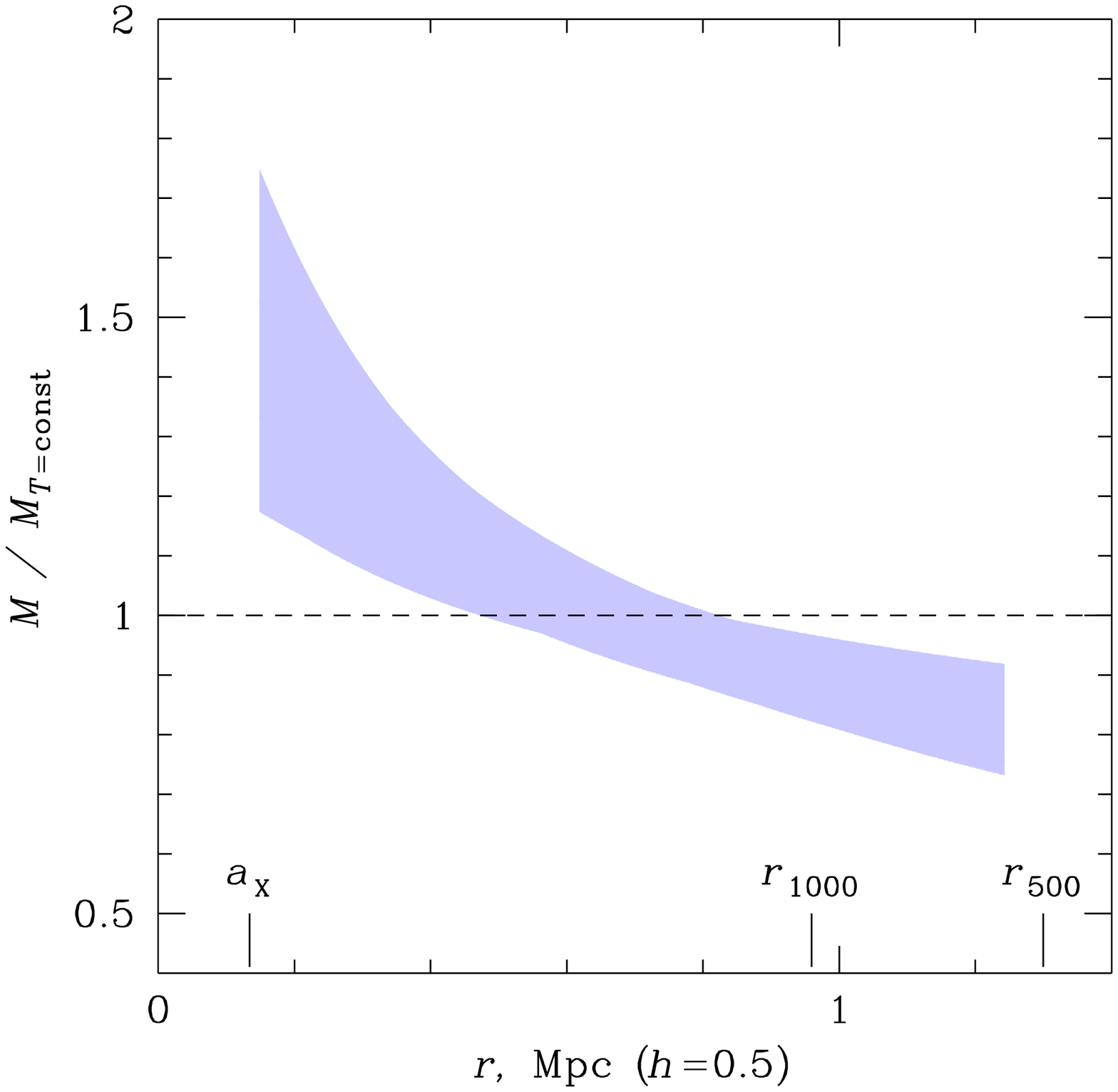}}

\rput[tl]{0}(9.5,20.7){\epsfxsize=8.5cm
\epsffile{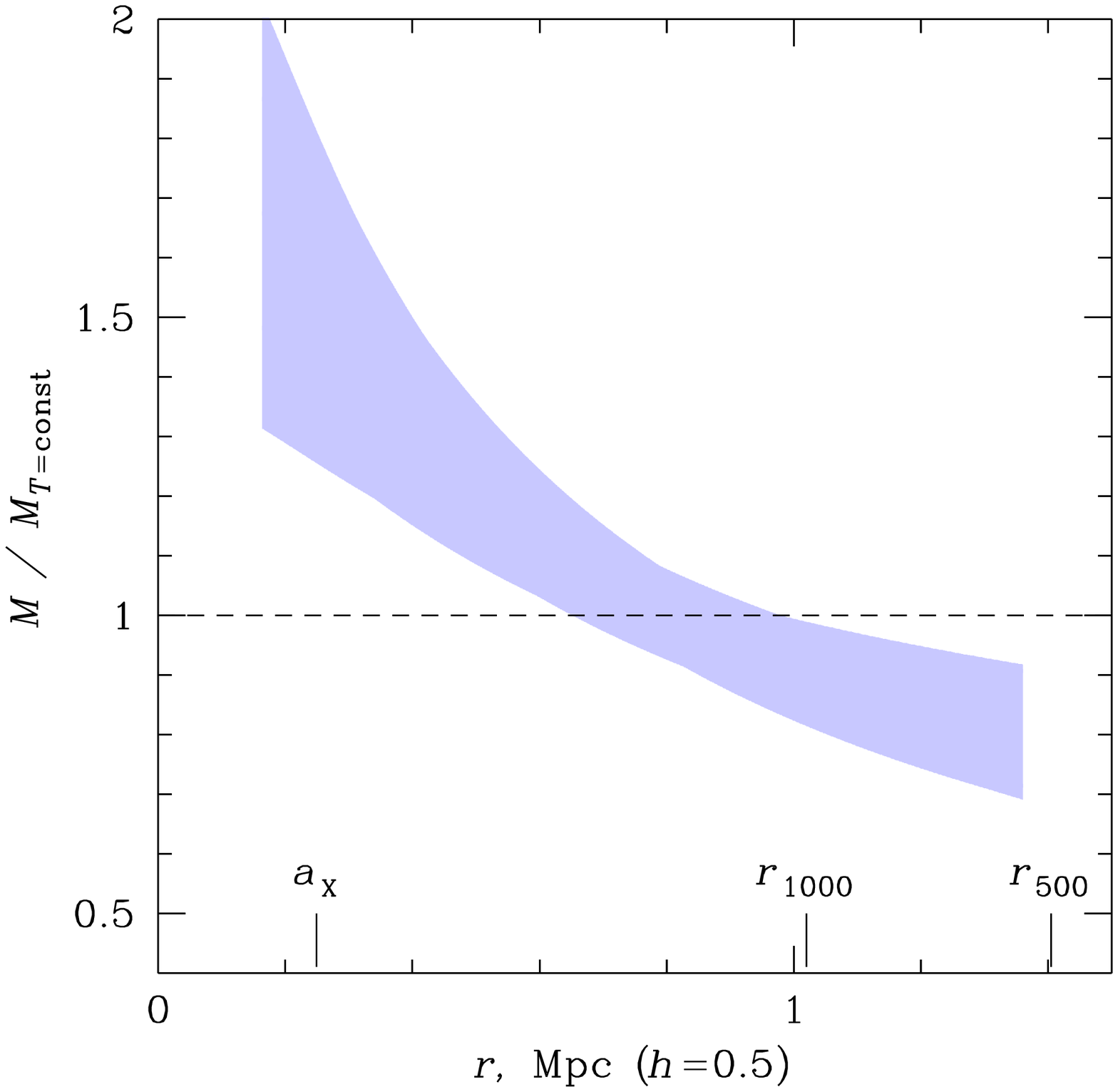}}

\rput[bl]{0}( 6.9,19.4){\small A2199}
\rput[bl]{0}(16.0,19.4){\small A496}

\rput[tl]{0}(0,12.4){
\begin{minipage}{18cm}
\small\parindent=3.5mm
{\sc Fig.}~4.---Polytropic correction to the mass profiles derived under the
isothermality assumption (see \S\ref{masssec}). Gray bands correspond to
90\% temperature errors of the observed profiles shown in Fig.\ 2 (excluding
the central cooling flow bins). The constant temperature is taken to be
equal to the cooling flow-corrected values. X-ray core radii $a_x$ are from
the \rosat\ PSPC data (\S\ref{rossec}), and $r_{1000}$ and $r_{500}$ are
same as in Fig.\ 2.
\end{minipage}
}
\endpspicture
\end{figure*}

\begin{figure*}[tb]
\pspicture(0,3.1)(18.5,21)


\rput[tl]{0}(0.4,20.7){\epsfxsize=8.5cm
\epsffile{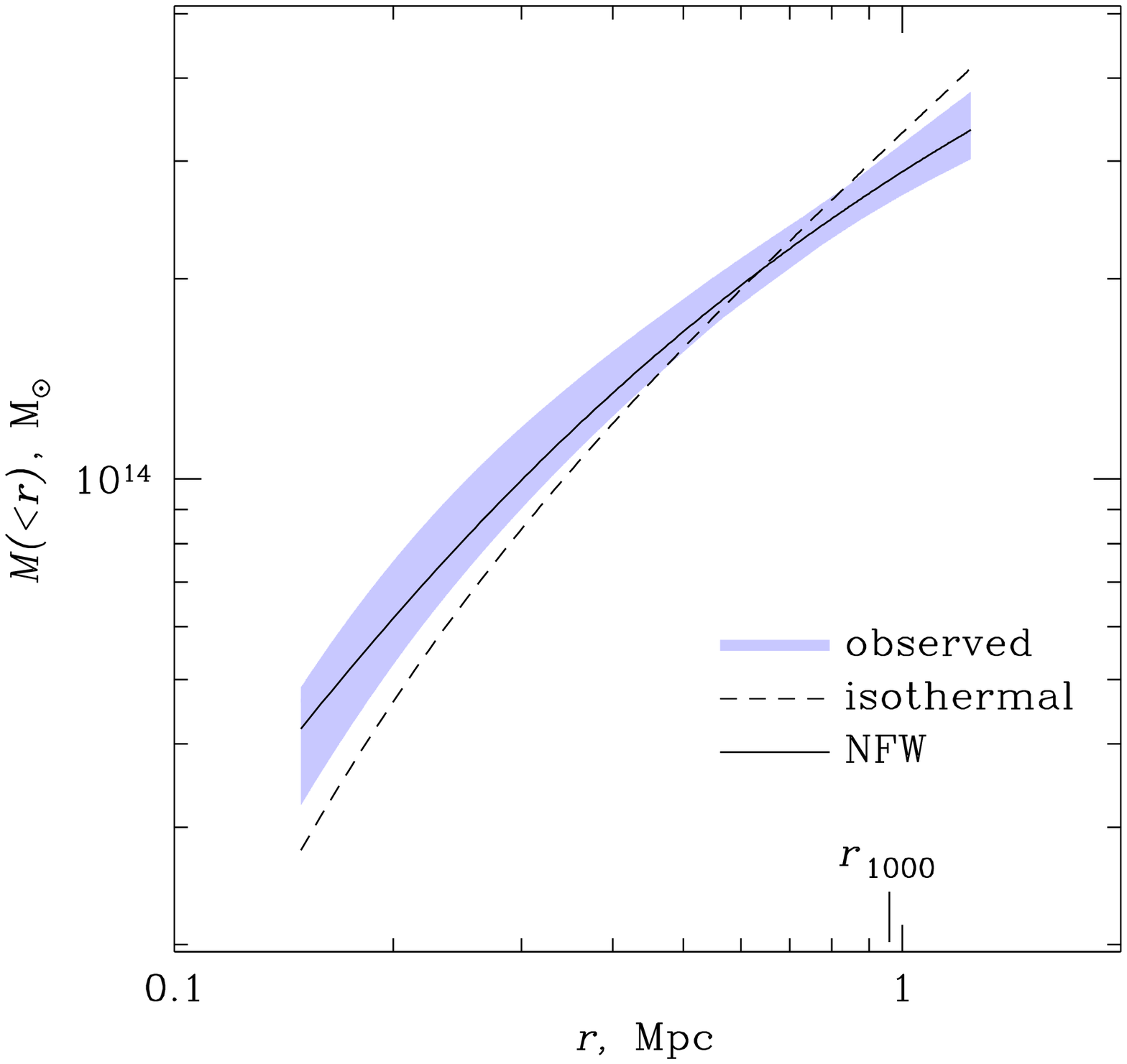}}

\rput[tl]{0}(9.5,20.7){\epsfxsize=8.5cm
\epsffile{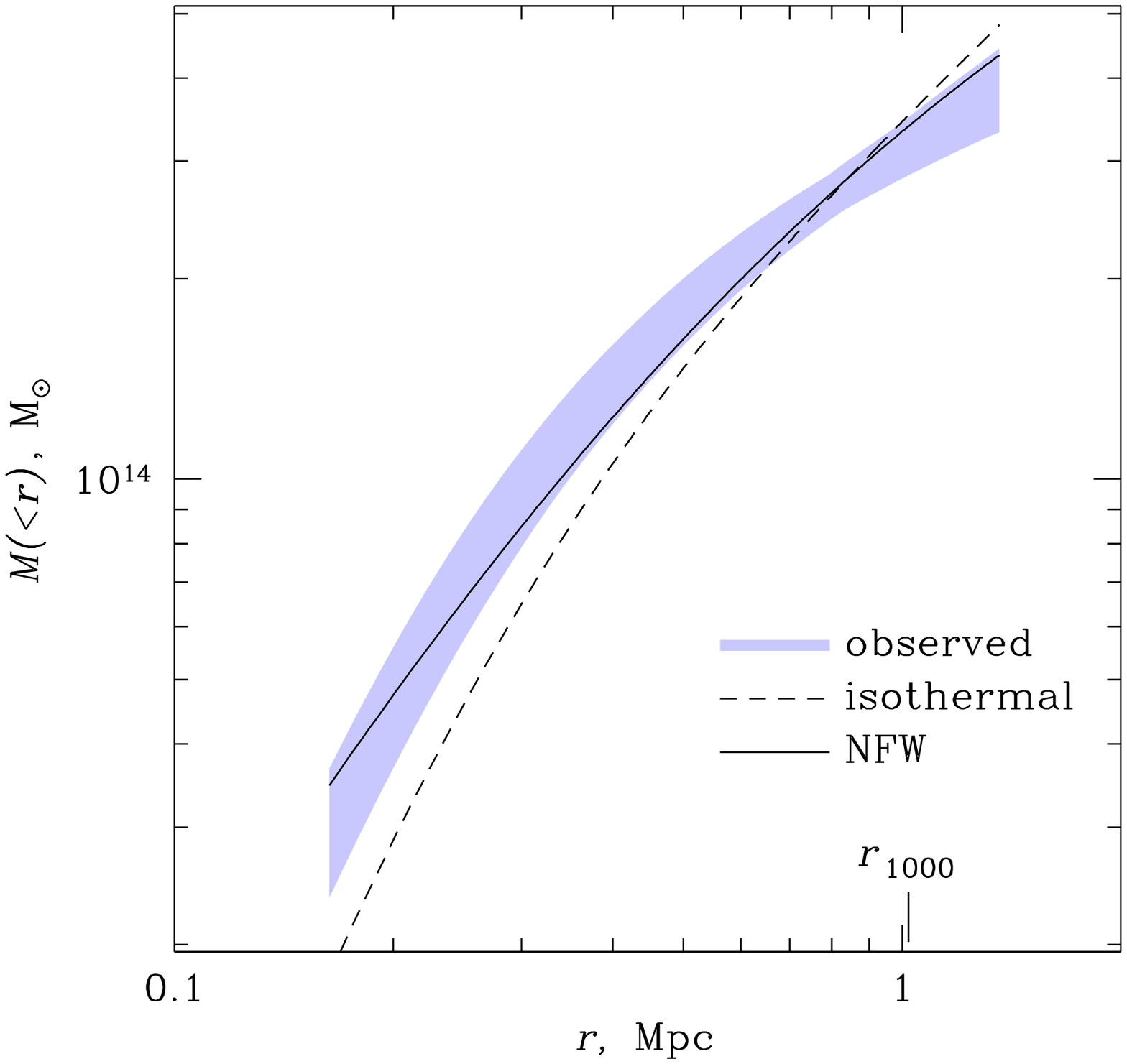}}

\rput[bl]{0}( 2.3,19.4){\small A2199}
\rput[bl]{0}(11.4,19.4){\small A496}


\rput[tl]{0}(0.4,12.9){\epsfxsize=8.5cm
\epsffile{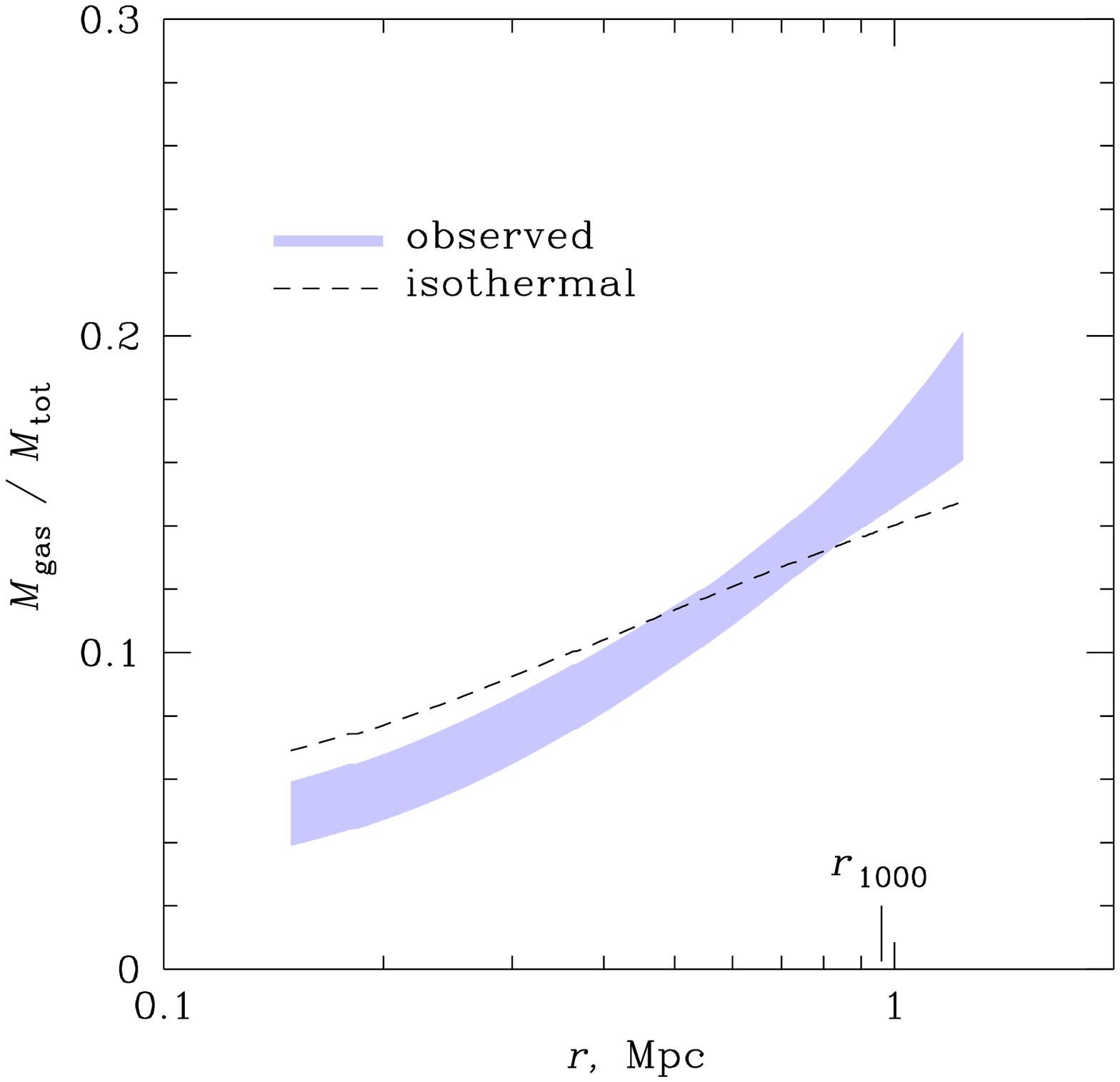}}

\rput[tl]{0}(9.5,12.9){\epsfxsize=8.5cm
\epsffile{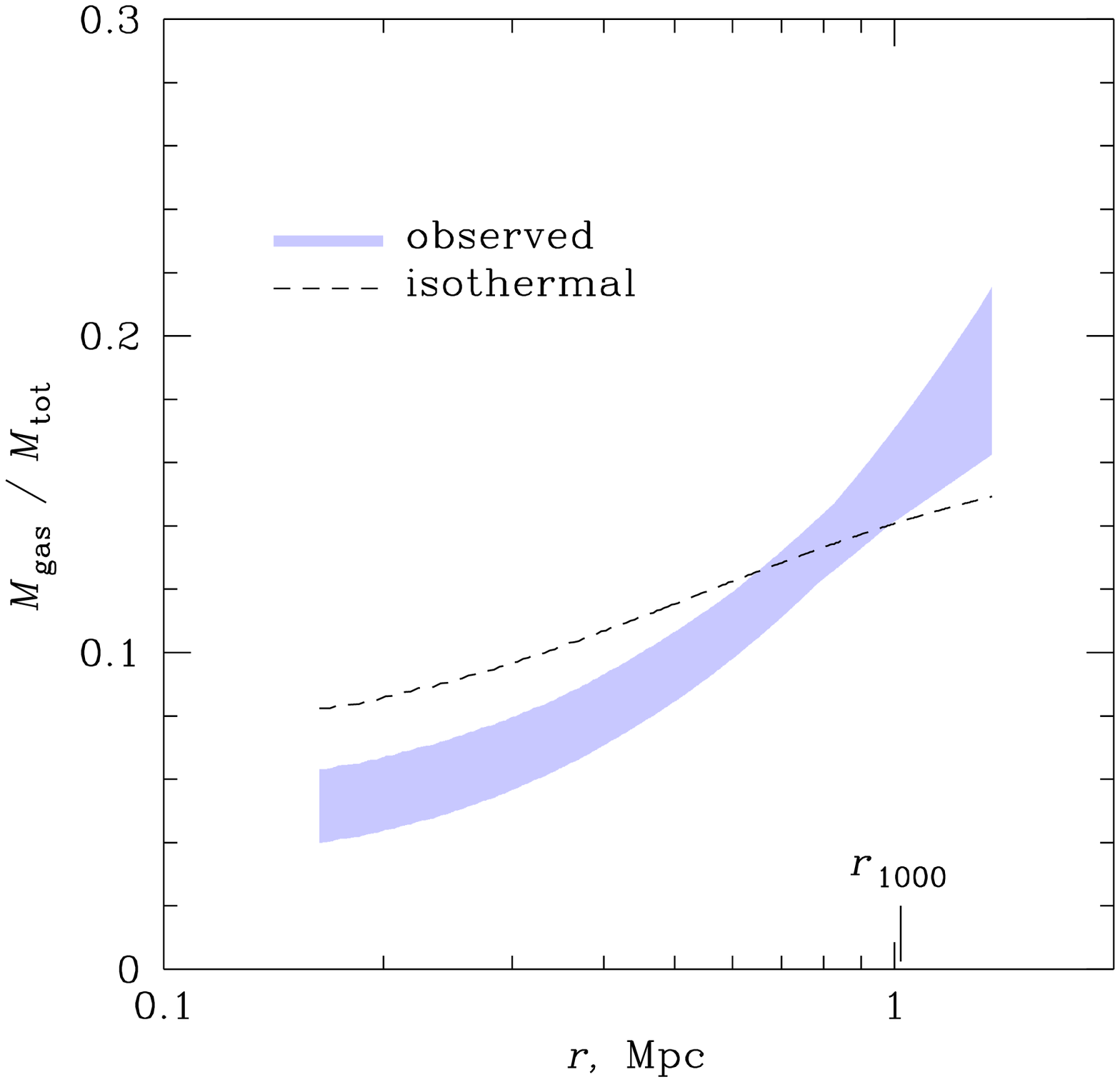}}

\rput[bl]{0}( 2.3,11.6){\small A2199}
\rput[bl]{0}(11.4,11.6){\small A496}

\rput[tl]{0}(0,4.6){
\begin{minipage}{18cm}
\small\parindent=3.5mm
{\sc Fig.}~5.---Total enclosed mass profiles and the corresponding gas
fraction profiles. All values are for $h=0.5$ (the total mass scales as
$h^{-1}$ and the gas fraction as $h^{-3/2}$). The 90\% confidence bands
(gray) correspond to those in Fig.\ 4. Best-fit profiles of the Navarro,
Frenk, \& White form are shown as solid lines in the mass panels. In all
panels, dashed lines show the profiles obtained assuming a constant
temperature (at its average, cooling flow-corrected value). The $r_{1000}$
values correspond to the best-fit NFW profiles shown here (see also Fig.\
6).
\end{minipage}
}
\endpspicture
\end{figure*}

\begin{figure*}[tb]
\pspicture(0,11.6)(18.5,21)

\rput[tl]{0}(0.4,20.7){\epsfxsize=8.5cm
\epsffile{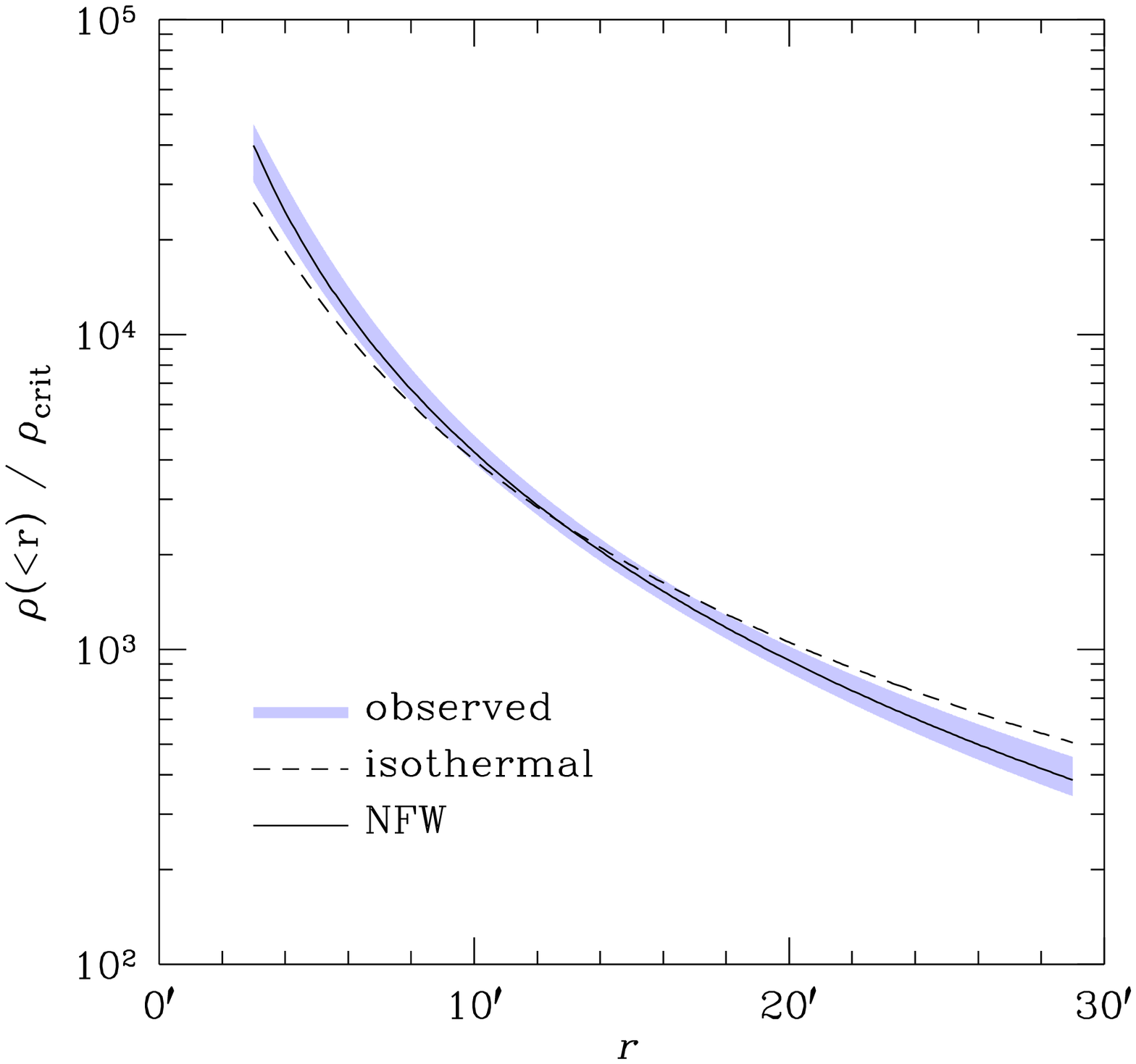}}

\rput[tl]{0}(9.5,20.7){\epsfxsize=8.5cm
\epsffile{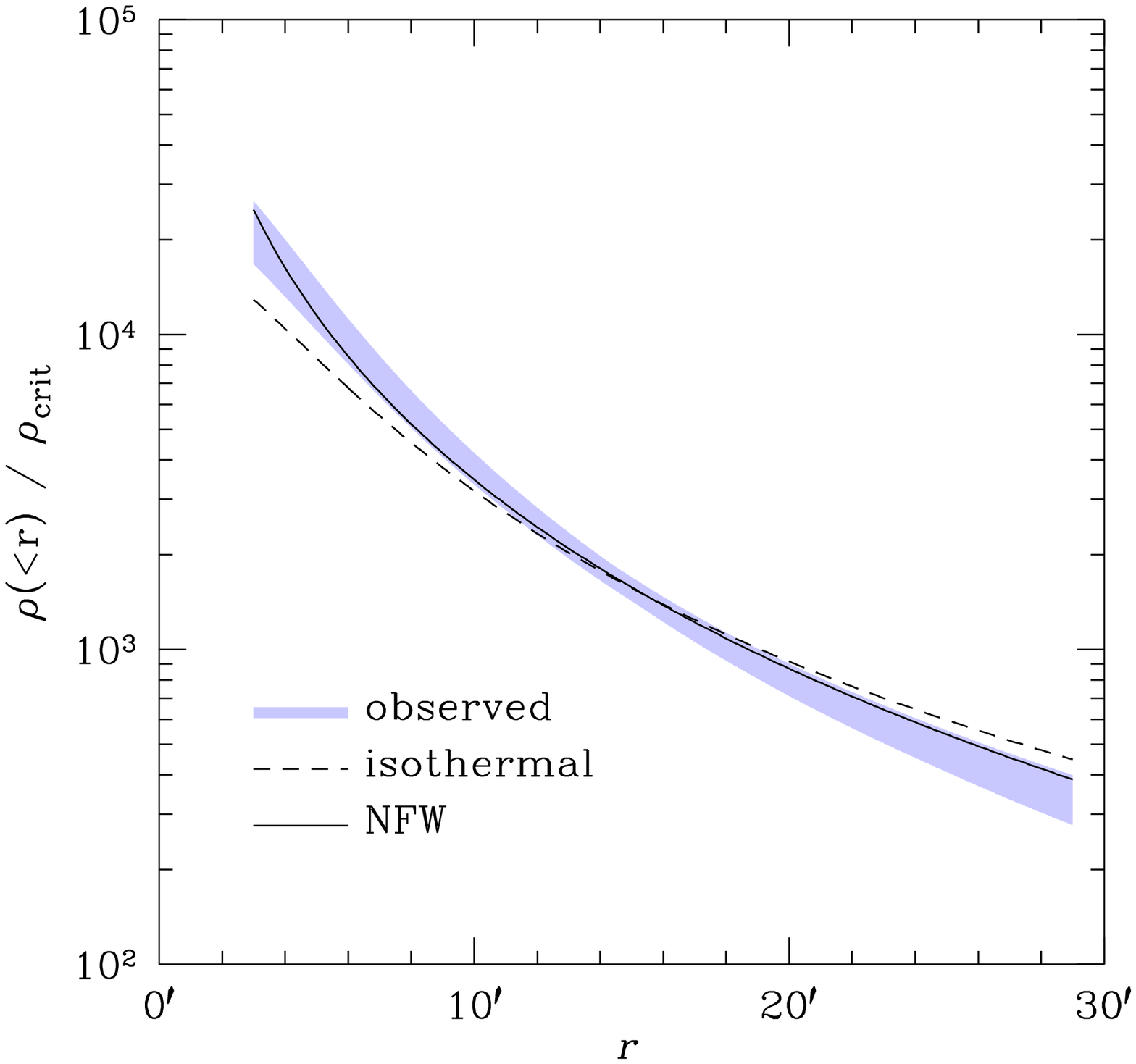}}

\rput[bl]{0}( 6.9,19.4){\small A2199}
\rput[bl]{0}(16.0,19.4){\small A496}

\rput[tl]{0}(0,12.4){
\begin{minipage}{18cm}
\small\parindent=3.5mm
{\sc Fig.}~6.---Average total overdensity within a given radius with respect
to the critical density at the clusters' redshifts, calculated for the
profiles shown in Fig.\ 5. The overdensity at a given angular distance does
not depend on $h$. The \asca\ data covers $r<25'$.
\end{minipage}
}
\endpspicture
\end{figure*}

\section{DISCUSSION}

\subsection{Uncertainty of Our Mass Values}

Although the polytropic model is a good representation of the observed
temperature profiles, it does not necessarily cover all possibilities that
could be consistent with the data. Since our best-fit model adequately
represents the temperature and its gradient, the best-fit mass values should
be unbiased (as long as the hydrostatic equilibrium assumption is valid).
However, at the extremes of the confidence intervals, some other acceptable
temperature models may result in slightly different mass profiles.
Therefore, our confidence intervals on mass may be underestimated. A more
exhaustive method of mass modeling would be to assume a certain functional
form of the dark matter profile and find parameter ranges consistent with
the data (e.g., Hughes 1989; Henry, Briel, \& Nulsen 1993; Loewenstein 1994;
MV; Nevalainen et al.\ 1999a,b). However, for the relatively high-quality
data on A2199 and A496, we have chosen a simpler approach with the
polytropic models, without giving undue importance to the formal error
estimates. Indeed, given the present accuracy of the temperature and density
profiles, the formal statistical uncertainties are already too small to be
physically meaningful (e.g., MV). Hydrodynamic simulations suggest that
systematic uncertainties in the method itself, such as the possible
deviations from spherical symmetry and hydrostatic equilibrium (in the form
of significant gas bulk motions), can give rise to {\em rms} mass errors of
about 15--30\% (e.g., Evrard, Metzler, \& Navarro 1996; Roettiger et al.\
1996; see a more detailed discussion in \S4.2 of MV). Those simulations
included merging clusters in the statistical sample, so the relaxed clusters
A2199 and A496 should have mass errors on the lower side of these estimates.

Another source of systematic mass uncertainty is the possible deviation of the
measured electron temperature from the local mean plasma temperature that
enters the hydrostatic equilibrium equation. Markevitch et al.\ (1996)
proposed such nonequality as a possible explanation for an unusually sharp
observed temperature gradient in A2163. Later theoretical work (Fox
\& Loeb 1997; Ettori \& Fabian 1998; Chi\`eze, Alimi, \& Teyssier 1998;
Takizawa 1998) concluded that for relaxed clusters, this effect should not
be significant within the radial distances presently accessible for accurate
X-ray temperature measurements (about half the virial radius).  Therefore,
it is safe to assume that the mass values within $r_{1000}$ obtained in this
paper are unaffected by this complication.

Finally, at these low redshifts, the unknown cluster peculiar velocity may
introduce a noticeable distance and mass error (e.g., a 1000 \kms\ velocity
would correspond to a 10\% error in the calculated mass). To summarize all
of the above, the true uncertainty of the masses of A2199 and A496 is
probably greater than our formal $\pm10$\% estimates and perhaps closer to
20--25\% (90\% confidence at $r_{1000}$), and is dominated by systematics.

\subsection{The ``Universal'' Mass Profile}

NFW have found that radial density profiles of equilibrium clusters in their
cosmological simulations can be approximated by a functional form
\mbox{$\rho(r) \propto (r/r_s)^{-1}(1+r/r_s)^{-2}$}. This form is a very
good description of our observed total mass profiles in the range of radii
covered by the temperature data, as shown in Fig.\ 5. Normalizations and
scale radii $r_s$ of the NFW profiles were selected to fit the observed mass
profiles.  For A2199, $r_s=0.18$ Mpc, and for A496, $r_s=0.36$ Mpc.%
\footnote{NFW included only dark matter in their simulations and it is
unclear whether the inclusion of gas would significantly change the shape of
the mass profiles. If we subtract the gas mass from our total mass profiles
(assuming, for example, the currently favored value of $h=0.65$), the
resulting dark matter profiles are well described by the same functional
form with slightly different parameter values.}
Extrapolating the best-fit NFW profiles to greater radii, we obtain the
NFW's concentration parameter $c\equiv r_{200}/r_s$ of about 10 and 6 for
the two clusters, respectively. According to the NFW's simulations, $c$ and
the total mass within $r_{200}$ are strongly correlated for a given
cosmological model. Our $c$ and $M_{200}$ values agree well with those for
several cosmological models considered by NFW, including CDM$\Lambda$ (for
that model, our observed masses correspond to $M_{200}/M_*\approx 5-6$ as
defined in NFW).  An isothermal profile would imply a less concentrated dark
matter distribution than that suggested by the NFW simulations (see also
Makino, Sasaki, \& Suto 1998).

The outer regions of other clusters for which relatively accurate mass
profiles were derived from the \asca\ temperature profiles (e.g., A2256, MV;
A3571, Nevalainen et al.\ 1999b) are consistent with the NFW profiles as
well, although the constraints are poorer. Thus, the similarity of the gas
temperature profiles found by MFSV for nearby relaxed clusters (outside the
cooling flow regions) appears to be due to the underlying ``universal'' dark
matter profile of the NFW form. Note that we do not consider cooling flow
regions due to their unknown temperature structure. As noted in MV and
Nevalainen et al.\ (1999a), in those few relaxed clusters without cooling
flows where the gas density profile exhibits a flat core all the way to the
center (e.g., A401), the dark matter cannot have an NFW central cusp because
the gas halo would be convectively unstable (see also Suto, Sasaki, \&
Makino 1998). On the other hand, it is likely that in clusters that do have
central dark matter cusps, the corresponding dip of the gravitational
potential causes the gas density peak and acts as a focus for a cooling
flow.

\subsection{Mass -- Temperature Scaling}

Our mass values within our $r_{1000}$ are a factor of $1.6-1.8$ below the
scaling relation between the total mass and emission-weighted average gas
temperature derived from the simulations by Evrard et al.\ (1996). The same
is true for other clusters (e.g., A2256, MV; A2029, Sarazin, Wise, \&
Markevitch 1998; A401, A3571, Nevalainen et al.\ 1999a,b). Note that
isothermal mass estimates are also lower than the Evrard et al.\ $M-T$
relation predicts. Given the agreement of our observed total (or dark) mass
profile with the ``universal'' profile from NFW as well as from Evrard et
al., the main source of this discrepancy apparently lies in the gas density
and temperature distributions. Indeed, as noted by MFSV, simulations predict
a less steep temperature decline than observed, and steeper gas density
profiles than observed (e.g., Vikhlinin et al.\ 1999). Both these effects
have the sign needed to cause the $M-T$ discrepancy. Because most of the
cluster X-ray emission originates in the central region, simulations that do
not sufficiently resolve the cluster core may predict an incorrect
(apparently too low) emission-weighted gas temperature. For example, the
Evrard et al.\ (1996) simulations have a resolution of 0.2--0.3 Mpc,
comparable to a typical cluster core radius. On the other hand, the
hydrostatic mass measurement can underestimate the true total mass if, for
example, there is significant gas turbulence. Indeed, simulations suggest
that there may be residual turbulence even in an apparently relaxed cluster
(e.g., Evrard et al.\ 1996; Norman \& Bryan 1998) resulting in a 10--15\%
underestimate of the mass within $r\sim r_{1000}$.

\subsection{Implications for the X-ray--Lensing Mass Discrepancy}

A2199 and A496 are representative examples of relaxed clusters and the
difference in their X-ray mass estimates using the measured temperature
profile from the isothermal estimates, shown in Fig.\ 4, generally applies
to other such clusters (see MV and MFSV). The upward revision of the mass
estimate in the inner part has one important implication, the convergence of
the X-ray and gravitational lensing mass estimates in the cluster central
regions. The strong lensing mass values (that usually correspond to $r\lax
0.2$ Mpc) often exceed by a factor of 2--3 the X-ray estimates made under
the assumptions of isothermality and a typical $\beta$-model density profile
(e.g., Loeb \& Mao 1994; Miralda-Escud\'e \& Babul 1995; Tyson \& Fischer
1995). In many cases, the lensing analysis is likely to overestimate the
mass as a result of substructure or projection (e.g., Bartelmann 1995). On
the X-ray side, some clusters are undergoing mergers and the hydrostatic
equilibrium method may give a wrong mass value. For those clusters which are
relaxed, the low-resolution X-ray image analysis may underestimate the gas
density gradient at small radii typical of the lensing measurements and may
be responsible for part of the disagreement (e.g., Markevitch 1997; Allen
1998). If the cluster has a strong cooling flow, the overall temperature can
be significantly underestimated if no allowance for the cool component is
made (Allen 1998), although for most clusters this correction is within
$\sim 20$\% (MFSV). Still, in many cases these effects alone are not
sufficient to account for the mass discrepancy.  It has been suggested,
e.g., by Miralda-Escud\'e \& Babul (1995) that a gas temperature gradient
could explain the discrepancy for the distant non-cooling-flow cluster
A2218. A temperature decline with radius has indeed been observed in A2218
by Loewenstein (1997) and Cannon, Ponman, \& Hobbs (1999) using \asca, while
MFSV find that such a declining profile is common among nearby clusters.

The analysis in \S\ref{masssec} has shown that within the core radius, the
commonly observed temperature gradient implies a mass that is higher than
the isothermal estimate by a factor of $\gax 1.5$. The reference isothermal
estimate uses a cooling flow-corrected temperature, so this effect is in
addition to the cooling flow-related mass correction of Allen (1998). If a
similar temperature gradient is common in more distant clusters, this
effect, together with others mentioned above, effectively resolves the mass
discrepancy. This seems to obviate the need for more exotic causes, such as
a significant magnetic field pressure or strong turbulence within cluster
cores (Loeb \& Mao 1994).

\subsection{Gas Fraction}

The lower panels in Fig.\ 5 show the gas mass fraction a function of radius
for the two clusters. At $r_{1000}$, we obtain similar values of $f_{\rm
gas}=0.161\pm 0.014$ and $0.158\pm 0.017$ for the two clusters,
respectively. These values are consistent with those for A2256, $0.14\pm
0.01$ at $r_{1000}$, obtained by MV using an \asca\ temperature profile, and
for A401 ($0.18^{+0.02}_{-0.04}$) and A3571 ($0.16^{+0.03}_{-0.01}$) from
Nevalainen et al.\ (1999a,b). Our values are also similar to the median
values for large samples of clusters analyzed using the isothermal
assumption: $f_{\rm gas}=0.168$ from Ettori \& Fabian (1999, scaled to
$r_{1000}$), and $f_{\rm gas}=0.160$ from Mohr, Mathiesen, \& Evrard (1999,
for clusters cooler than 5 keV). The latter similarity is due to the fact
that the effect of the radial temperature decline on mass is small at this
radius; at greater radii, the isothermal analysis underestimates the gas
fraction as Fig.\ 5 shows.

The values of the cluster gas fraction from X-ray analysis are often used to
place constraints on the cosmological density parameter ($\Omega_0\lax
0.3$), under the assumption that $f_{\rm gas}$ in clusters is representative
of the Universe as a whole (White et al.\ 1993 and many later works).
However, $f_{\rm gas}$ increases with radius even if one assumes a constant
gas temperature (e.g., David, Jones, \& Forman 1995; Ettori \& Fabian 1999),
and the true increase is steeper as seen in clusters with measured
temperature profiles. Fig.\ 5 shows that $f_{\rm gas}$ increases by a factor
of 3 between the X-ray core radius and $r_{1000}$ and does not show any
evidence of flattening at large radii and hence asymptotically reaching a
universal value. Although at some radius, the cluster must merge
continuously into the infalling matter with a cosmic mix of components, that
presumably happens at the infall shock radius of $\sim 2-3\, r_{1000}$, well
beyond the region presently accessible to accurate measurements. Note that
both the dark matter and the gas mass within a given radius, and thus
$f_{\rm gas}$, are dominated by the contribution at large radii.
Cosmological simulations suggest that at smaller radii, a deviation from the
universal $f_{\rm gas}$ value is not large (e.g., Frenk et al.\ 1996).
However, at the present stage, the simulations do not accurately reproduce
the observed gas density, temperature and $f_{\rm gas}$ profiles (e.g.,
MFSV; Vikhlinin et al.\ 1999). We conclude that our $f_{\rm gas}$ values for
A2199 and A496 are consistent with the constraints on $\Omega_0$ derived in
earlier works (e.g., Ettori \& Fabian 1999; Mohr et al.\ 1999), but caution
that such estimates at present involve a large extrapolation. Future
observatories \chandra\ and \xmm\ will be capable of studying the cluster
outermost regions and possibly determining the asymptotic value of $f_{\rm
gas}$.

\section{SUMMARY}

The \asca\ gas temperature maps and radial profiles for A2199 and A496
indicate that these systems are representative examples of relaxed,
moderately massive clusters. Our high quality temperature data imply total
mass profiles that are in good agreement with the NFW simulated
``universal'' profile over the range of radii covered by the data
($0.1\,{\rm Mpc}< r < r_{1000}\approx 1\,{\rm Mpc}$). Because the
temperature profiles of these two clusters are similar to the average
profile for a large sample of nearby clusters in MFSV, this agreement
indicates that the NFW profile is indeed common in nearby clusters. The
upward revision of the total mass at small radii, by a factor of $\gax 1.5$
compared to an isothermal analysis, may reconcile X-ray and strong lensing
mass estimates in distant clusters. The observed mass profile also implies a
gas mass fraction profile steeply rising with radius. While our $f_{\rm
gas}$ values at $r_{1000}$ support earlier upper limits on $\Omega_0$, the
steep increase of $f_{\rm gas}$ with radius, not anticipated by most cluster
simulations, suggests that we may not yet have correctly determined the
universal baryon fraction and caution is needed in such analysis.

\acknowledgments

This work was supported by NASA contracts and grants NAS8-39073,
NAG5-3057, NAG5-4516, NAG5-8390,
and by the Smithsonian Institution.

\appendix
\section{PROJECTION OF THE POLYTROPIC GAS TEMPERATURE PROFILE}

Assuming a spherically symmetric gas density distribution of the form
$\rho(r)\propto (1+r^2/a^2)^{-\frac{3}{2}\beta}$ (and taking $a=1$ for
clarity) and the polytropic temperature profile
\begin{equation}
\label{tpoly}
T(r)\propto \rho^{\gamma-1}(r),
\end{equation}
one can calculate a temperature profile that is emission-weighted
(by $\rho^2$)
along the line of sight $l$, as a function of the projected distance from
the center $x$ (such that $r^2=x^2+l^2$), as
\begin{equation}
T_{\rm proj}(x)=
\frac{\int_0^{\infty}T(r)\,\rho^2(r)\,dl}{\int_0^{\infty}\rho^2(r)\,dl}\propto
\frac{\int_0^{\infty} \rho^{1+\gamma}(r)\, dl}{\int_0^{\infty}\rho^2(r)\, dl}
\propto
\frac{\int_0^{\infty} (1+x^2+l^2)^{-\frac{3}{2}\beta(1+\gamma)}\, dl}
{\int_0^{\infty}      (1+x^2+l^2)^{-3\beta}\, dl}
\propto
\frac{(1+x^2)^{-\frac{3}{2}\beta(1+\gamma)+\frac{1}{2}}}
     {(1+x^2)^{-3\beta+\frac{1}{2}}}
=(1+x^2)^{-\frac{3}{2}\beta(\gamma-1)}
\propto T(x).
\end{equation}
That is, the resulting projected temperature profile has the same shape as
the real (three-dimensional) profile in eq.\ (\ref{tpoly}). The relative
normalization of the projected profile at $x=0$ (and, therefore, at all
radii) can easily be derived using the above formulae and is found to be:
\begin{equation}
\label{norm}
\frac{T_{\rm proj}}{T}=
\frac{\Gamma\left[\frac{3}{2}\beta(1+\gamma)-\frac{1}{2}\right]\,\,
\Gamma(3\beta)}
{\Gamma\left[\frac{3}{2}\beta(1+\gamma)\right]\,\,\Gamma(3\beta-\frac{1}{2})}
.
\end{equation}
The normalization of the projected temperature profile is slightly smaller
than that of the three-dimensional profile.  For $\beta>0.5$ and
$\gamma<5/3$, their difference is less than $\sim 20$\%.

Strictly speaking, different temperatures along the line of sight are not
simply weighted with $\rho^2$; to obtain an exact projected temperature, a
single-temperature fit to a multi-temperature spectrum should be performed
in the \asca\ energy band. However, the above similarity of the projected
and real profiles holds for any weighting that is proportional to $\rho^2
T^\alpha$, which approximates a wide range of possibilities. The
normalization (\ref{norm}) changes only weakly if $\alpha\neq 0$.  For
example, taking $\alpha=0.5$ (weighting with a bolometric emissivity)
instead of $\alpha=0$ changes the normalization for our clusters by only
1\%, and by less than 5\% for any reasonable $\beta$ and $\gamma$. We have
therefore assumed $\alpha=0$ in \S\ref{tprof} for simplicity.

\end{document}